\documentstyle[12pt]{article}
\begin{document}
\title{\vspace{-15mm}
       {\normalsize \hfill
       \begin{tabbing}
       \`\begin{tabular}{l}
	 hep--th/9407198 \\
	 RI--8--94  \\
	 July 1994 \\
	\end{tabular}
       \end{tabbing} }
       \vspace{8mm}
\setcounter{footnote}{1}
A truly marginal deformation of SL(2,R) in a null direction}

\renewcommand{\thefootnote}{\fnsymbol{footnote}}
\vspace{5mm}

\author{
\setcounter{footnote}{2}
Stefan F\"orste
\thanks{e-mail: sforste@vms.huji.ac.il, Work supported by a grant
of MINERVA}\\ {\normalsize \em Racah Institute of Physics} \\
{\normalsize \em The Hebrew University, 91904 Jerusalem, Israel}}
\date{}

\maketitle

\begin{abstract}
We perform a marginal deformation of the $SL(2,R)$ WZW model in a null
direction. If we send the deformation parameter to infinity we obtain
a linear dilaton background plus two free bosons. We show in addition
that such a background can be obtained by a duality transformation of the
undeformed WZW model. In the end we indicate how to generalize the
given procedure.
\end{abstract}

\renewcommand{\arraystretch}{2.0}
\renewcommand{\thefootnote}{\alph{footnote}}
\newcommand{\be}{\begin{equation}}
\newcommand{\ee}{\end{equation}}
\newcommand{\ba}{\begin{array}}
\newcommand{\ea}{\end{array}}
\newcommand{\vsf}{\vspace{5mm}}
\newcommand{\NP}[3]{{\em Nucl. Phys.}{ \bf B#1#2#3}}
\newcommand{\marpar}{\marginpar[!!!]{!!!}}

One interesting observation in string theory is that theories with a completely
different sigma model interpretation belong actually to the same conformal
field theory. The reason for that is the existence of certain symmetries
which change the background field configuration in a non trivial way.
These symmetries are e.g.\ duality transformations \cite{buscher} and
$O(d,d,R)$
\cite{venez} transformations.
Duality \cite{x} and the discrete subgroup $O(d,d,Z)$ \cite{amit} connect sigma
models belonging to the same conformal field theory whereas the $O(d,d,R)$
transforms between conformal backgrounds not necessarily corresponding to the
same conformal field theory. (A review about these symmetries is given in
\cite{gpr}.)
Another interesting issue is that two different sigma models might be connected
by a line of a truly marginal deformation \cite{sen,giv}.
A way to find a truly marginal deformed model is to take a
conformally invariant one parameter family of actions $S(\alpha)$ with a
chiral and anti chiral symmetry. Then one has to show that a variation in
$\alpha $ corresponds to a perturbation by a product of the chiral
and anti chiral current, i.e.\ that the differential equation
\begin{equation}    \label{mpf}
\frac{\partial S(\alpha )}{\partial \alpha } \sim \int J(\alpha )
\bar{J}(\alpha )
\end{equation}
holds. Moreover one has
to check that for $\alpha $ goes to zero one arrives at the original theory
infinitisimaly perturbed by a truly marginal operator.
In \cite{giv} it was shown that deforming a $WZW$ model of a group $G$ leads
in one end point of the deformation to a direct product of the gauged model
$G/U(1)$ and a non compact $U(1)$, whereas in the other end $G/U(1)$ is
replaced by it's dual.
One can use the deformation of
the $SU(2)$ WZW model given in \cite{sen,giv} in order to obtain a truly
marginal deformed $SL(2,R)$ WZW model. There is an
interesting difference between those two models. Deforming the $SL(2,R)$ in
the $J_2 \bar{J}_2$ direction one gets at one end of the deformation line the
2d euclidean black hole $\times$ non compact
$U(1)$ whereas
deforming in the
$J_3 \bar{J}_3$ direction gives in the end the 2d lorentzian black hole
$\times$
non compact $U(1)$ \cite{efr}. In the other end the corresponding dual black
hole backgrounds $\times$ non compact $U(1)$ appear. ($J_2$ and $J_3$ are the
$i\sigma_2$ and $\sigma_3$ components of the
Kac-Moody current, respectively.) So, there is a difference between the
deformations into a space like or time like direction. In the present paper we
are
going to consider
a third possibility namely the deformation in a null direction. \\
We will give a deformed model satisfying (\ref{mpf}) and the initial
condition. As a check we will show that the deformed model is a conformal field
theory to one loop order\footnote{This is actually not necessary since a truly
marginal deformed exact CFT gives an exact CFT.}.
Furthermore we will see
that at the end points ($\alpha \rightarrow \pm \infty $) we get the dual
of our original model.\\
\mbox{}\\
We parametrize the $SL(2,R)$ as follows
\begin{equation}
g=g_- g_0 g_+,
\end{equation}
with
\begin{equation}
g_-=\left(
\begin{array}{c c}
1 & 0 \\ \frac{x_-}{\sqrt{2}} & 1 \end{array}\right) \;\;\;\;\;\;
g_0 =\left(\begin{array}{ c c} e^{\frac{x}{2}} & 0 \\ 0 & e^{-\frac{x}{2}}
\end{array}\right) \;\;\;\;\;\;
g_+ = \left( \begin{array}{c c} 1 & \frac{x_+}{\sqrt{2}} \\ 0 & 1
\end{array}\right).
\end{equation}
Then we obtain for the WZW model
\begin{equation}
S_{WZW}=\frac{k}{16\pi}\int tr\left( g^{-1}dg\wedge *g^{-1}dg \right)
+\frac{ik}{24\pi}\int tr\left( g^{-1}dg^3\right)
\end{equation}
the following action\footnote{The euclidean version can be
obtained by taking $x_+$ and $x_-$ as complex conjugated variables.}
\begin{equation}      \label{wzw}
S_{WZW}=\frac{k}{4\pi}\int d^2 z \left( \frac{1}{2}\partial x \bar{\partial} x
+ e^x \partial x_+ \bar{\partial} x_- \right),
\end{equation}
with the conserved currents
\begin{equation}
J \sim e^x \partial x_+ \; , \;\;\;\;\;  \bar{J} \sim e^x \bar{\partial}
x_- \; .
\end{equation}
Now we will show that the one parameter family of theories
\begin{equation}      \label{dwzw}
S(\alpha ) = \frac{k}{4\pi}\int d^2 z\left( \frac{1}{2} \partial x
\bar{\partial} x + \frac{e^x}{1-\alpha e^x } \partial x_+
\bar{\partial} x_- \right)
\end{equation}
is the integrated marginal deformation of (\ref{wzw}) with respect to
$J \bar{J}$. (The action (\ref{dwzw}) can not be obtained by using the known
$SU(2)$ deformation \cite{sen,giv}. It can rather be guessed, or a systematic
way is to start with an infinitisimal perturbation resulting in an
infinitisimal change in the currents. Repeating that procedure will result in a
geometrical series which sums up to (\ref{dwzw}).)\\
{}From (\ref{dwzw}) we obtain the chiral currents
\begin{equation}    \label{cur}
J(\alpha ) \sim \frac{e^x}{1 -\alpha e^x}\partial x_{+} \; ,
\;\;\;\;\;\; \bar{J}(\alpha ) \sim \frac{e^x}{1-\alpha e^x}\bar{\partial}x_- .
\end{equation}
Differentiating  $S(\alpha)$ with respect to $\alpha$ provides
\begin{equation}   \label{proof}
\frac{\partial S(\alpha)}{\partial \alpha} \sim \int d^2z J(\alpha )
\bar{J}(\alpha ).
\end{equation}
For $\alpha = 0$ we obtain the original model. In the original theory the
measures for the functional integral are defined with respect
to the undeformed background. Introducing measures referring to
the deformed background results in a jacobian which can be expressed
by an additional dilaton term \cite{buscher}. Taking that into account we
obtain
finally\footnote{In \cite{ark} this background was obtained as a gauged
$SL(2,R)\times R/R$ WZW model.}
\begin{equation} \label{ecft}
S(\alpha ) = \frac{k}{4\pi }\int d^2z \left( \partial x
\bar{\partial} x + \frac{e^x}{1-\alpha e^x}\partial x_+ \bar{\partial}
x_-\right) -\frac{1}{8\pi}\int d^2z\sqrt{g}R\log(1-\alpha e^x ).
\end{equation}
Since the dilaton does not contain $x_{\pm}$ it does not change the
definition of the currents (\ref{cur}). It's variation with $\alpha $
cancels against the variation of the measure \cite{giv}. Hence
(\ref{proof}) still holds.
Modifying the consideration given in the appendix of \cite{giv} slightly
one can convince himself that (\ref{ecft}) is the only solution satisfying
(\ref{mpf}) and the initial condition.
As a test we checked the
one loop conformal invariance. We found that (\ref{ecft}) is really a
conformal field theory with central charge
\begin{equation}
c = 3 + \frac{12}{k} + o\left( \frac{1}{k^2}\right)
\end{equation}
at one loop level. The central charge does not depend on $\alpha $ and
coincides with the central charge of the original model \cite{kni}
\begin{equation}
c=\frac{3k}{4-k}.
\end{equation}
The deformation parameter $\alpha $ can run from minus infinity to
plus infinity giving $SL(2,R)$ at zero. In order to check what happens at
the borders we set $\alpha = \frac{1}{\epsilon}$ rescale
$$x_{\pm} \rightarrow \frac{1}{\sqrt{\epsilon}} x_{\pm}  $$
and send $\epsilon$ to zero. That results in
\begin{equation}     \label{limes}
S(\infty ) = \frac{k}{4\pi }\int d^2z \left( \frac{1}{2}\partial x
\bar{\partial}x -\partial{x_+}\bar{\partial}x_-\right)
-\frac{1}{8\pi}\int d^2z \sqrt{g}R x.
\end{equation}
This is a theory consisting of a linear dilaton background and
two free bosons. (At minus infinity we get a plus sign in front of $\partial
x_{+} \bar{\partial} x_{-}$.) The linear dilaton background appears also when
we
gauge the $SL(2,R)$ with respect to $\bar{J}$ from the left and with respect to
$J$
from the right \cite{klim}. So, at the end point of the deformation the
deformed
theory decouples into a direct product of the gauged theory and two
bosons. This is quite similar to the case discussed in \cite{sen,giv}
where at the end points of the deformation direct products occur which
contain the axial or vector gauged WZW model.
\\ \mbox{ }
\\
Now we are going to show that $S(0)$
and $S(\infty )$ describe
equivalent quantum theories. As a first step we gauge
the translational symmetry of $x_-$ in (\ref{wzw}) and write down the
identity
\begin{equation}      \label{id}
\int {\cal D}x_+ {\cal D}x_- e^{-S_{WZW}}=
\int {\cal D} A {\cal D} \bar{A}{\cal D} \lambda {\cal D}x_+
{\cal D}x_- \det(\partial \bar{\partial}) e^{-S(A,\bar{A})},
\end{equation}
where
\begin{equation}
S(A,\bar{A}) =\frac{k}{4\pi}\int d^2z\left\{ \frac{1}{2} \partial x
\bar{\partial} x + e^x\partial x_+ \bar{\partial}x_- +\bar{A}
e^x \partial x_+ + \lambda \left( \partial \bar{A} -
\bar{\partial} A \right) \right\}.
\end{equation}
The identity (\ref{id}) is known from duality transformations
\cite{x}. One can check (\ref{id}) by first performing the
$\lambda$ integration and then using the resulting $\delta$-function
to express the gauge fields as a pure gauge, (the determinant
$\det (\partial \bar{\partial} )$ cancels when one performs a
field redefinition in such a way that there are no derivatives in
the argument of the $\delta$-function). Being a pure gauge field
$\bar{A}$ can be absorbed in $\bar{\partial}x_{-}$. In order to
obtain a different background we do not perform the $\lambda $
integration, instead we integrate out $\bar{A}$ which provides a
$\delta$-function,
$$\delta \left( e^x\partial x_+ - \partial \lambda\right) .$$
We use that $\delta$-function to integrate out $x_+$ and get
\begin{equation}    \label{id1}
\int{\cal D}x_+{\cal D}x_- e^{-S_{WZW}} = \int {\cal D}A{\cal D}\lambda
{\cal D}x_- \det(\bar{\partial})e^{-S_1},
\end{equation}
where $S_1$ is given by
\begin{equation}     \label{s1}
S_1 = \frac{k}{4\pi}\int d^2z \left( \frac{1}{2}\partial x
\bar{\partial}x +\partial \lambda \bar{\partial}x_- +
A\bar{\partial}\lambda \right) -\frac{1}{8\pi}\int d^2z
\sqrt{g}Rx.
\end{equation}
The linear dilaton in (\ref{s1}) arises from the determinant
$$\left( \det  e^x \partial \right)^{-1} =\left( \det e^x
\det\partial\right)^{-1}  . $$
Now integrating over $\lambda$ gives
$$\delta \left( \bar{\partial}A+ \partial \bar{\partial}x_- \right). $$
Again we perform a field redefinition in such a way that there are
no derivatives in the $\delta$-function. Hence the $A,x_-$ integrations
lead just to
$$ \det\bar{\partial}^{-1} \left( \det\bar{\partial}\partial\right)^{-1}. $$
The first factor cancels with $\det \bar{\partial}$ in (\ref{id1})
and the second factor can be expressed as a functional integral
of two free bosons. Hence we get finally
\begin{equation}
\int {\cal D}x_+{\cal D}x_-e^{-S_{WZW}} = \int {\cal D}y_+
{\cal D}y_- e^{-\hat{S}},
\end{equation}
with
\begin{equation}  \label{du}
\hat{S} = \frac{k}{4\pi}\int d^2z \frac{1}{2}\partial x \bar{\partial} x
-\frac{1}{8\pi}\int d^2z \sqrt{g}Rx +\frac{1}{4\pi}\int d^2 z
\partial y_+ \bar{\partial}y_-
\end{equation}
which is exactly the theory we obtained by an infinite marginal
deformation\footnote{Since we did not consider global effects when calculating
the dual action it might differ from $S(\infty)$ topologically.}.
So, the $\alpha =0$ and the $\alpha = \infty $ points of our family
of exact conformal field theories (\ref{ecft}) are connected by a
duality transformation. In fact, one can do this duality tranformation at any
finite $\alpha$ and gets always (\ref{du}). That shows that the whole
deformation line consists of equivalent conformal field theories up to global
issues which were not taken into account. However, global issues might be
important, e.g.\ neglecting them one gets the result that the conformal field
theory of a free boson compactified on a circle is dual to a non compact free
boson independent of the radius of the circle. (The more familiar duality
transformation replacing
the metric by it's inverse can be obtained by gauging the translational
invariance of $x_{\pm}$ simultaneously.)\\
\mbox{ }\\
So, in difference to the $SU(2)$ WZW model the $SL(2,R)$ WZW model allows
(at least) for three different truly marginal deformations. In the end points
we get either a direct product of the 2d lorentzian black hole and a free
boson, the 2d euclidean black hole and a free boson, or a linear dilaton
background and two free bosons. That shows that the $c=1$ non critical string
and 2d black hole physics are connected (cf.\cite{dijk}) though not equivalent
(since it is not possible to rotate a null direction into a time or space like
direction).\\
\mbox{ }\\
Finally we remark that the presented considerations can be
generalized for any CFT of the form
\begin{equation}   \label{gen}
S  = \frac{k}{4\pi}\int d^2z\left( E_{\mu \nu}(r)\partial r^{\mu}
\bar{\partial}r^{\nu} + F(r)\partial x_+\bar{\partial}x_- \right)
+\frac{1}{4\pi}\int d^2 z \sqrt{g}R\Phi (r).
\end{equation}
The deformed model is\footnote{That this deformation respects the conformal
invariance is shown in \cite{ark} in a different way.}
\begin{equation}
\begin{array}{r c l}
S(\alpha )&=&\frac{k}{4\pi}\int d^2 z \left( E_{\mu \nu}(r)\partial r^{\mu}
\bar{\partial}r^{\nu}+\frac{F(r)}{1-\alpha F(r)}\partial x_+
\bar{\partial}x_- \right)\\& & +\frac{1}{4\pi}\int d^2z\sqrt{g}R\left( \Phi(r)-
\frac{1}{2} \log(1-\alpha F(r))\right).
\end{array}
\end{equation}
With
\begin{equation}
J(\alpha ) \sim \frac{F(r)}{1-\alpha F(r)}\partial x_+ \; , \;\;\;\;\;\;
\bar{J}(\alpha ) \sim \frac{F(r)}{1-\alpha F(r)} \bar{\partial} x_-
\end{equation}
it is easy to check that
\begin{equation}
\frac{\partial S}{\partial \alpha} \sim \int J(\alpha ) \bar{J} (\alpha ).
\end{equation}
Models of the form (\ref{gen}) have been considered in \cite{klim} in the
connection with gauging by null subgroups. A detailed application of
our marginal deformation to the more general case (\ref{gen}) will
be done soon. Forthcoming
work will also address problems left open here, e.g.\ global issues, or whether
the discussed three directions are the only non equivalent ones, (in that
context it might be interesting to study a deformation where for example the
chiral current belongs to a null direction and the anti chiral current belongs
to a space or time like direction).
\\ \mbox{ }\\
I would like to thank Amit Giveon, Eliezer Rabinovici and Gautam Sengupta for
very useful discussions. Moreover I would like to thank Arkady Tseytlin for
comments.

\end{document}